\documentclass{aastex}         
\usepackage{spr-astr-addons}   

\usepackage[cp866]{inputenc}
\usepackage[english]{babel}
\usepackage{graphicx}
\usepackage{amsmath}
\usepackage{amssymb}
\usepackage{epsf}
\usepackage{bm}

\def\la{\;\raise0.3ex\hbox{$<$\kern-0.75em\raise-1.1ex\hbox{$\sim$}}\;}
\def\ga{\;\raise0.3ex\hbox{$>$\kern-0.75em\raise-1.1ex\hbox{$\sim$}}\;}
                                                                                     
\newcommand{\emaila}{calisto@rbcmail.ru, akaurov@gmail.com,  kam@astro.ioffe.ru}

\shorttitle{Quasi-periodical features}
\shortauthors{Ryabinkov, Kaurov,  Kaminker}

\begin{document}

%
\title{Quasi-periodical features in the distribution   \\
of Luminous Red Galaxies}  


\author{A.~I.~Ryabinkov\altaffilmark{1}}  
\and  
\author{A.~A.~Kaurov\altaffilmark{2}} 
\and   
\author{A.~D.~Kaminker\altaffilmark{1,3}}  	 
\email{\emaila}

\altaffiltext{1}{Ioffe Physical-Technical Institute,
Politekhnicheskaya 26, 194021 St.~Petersburg, Russia.} 
\altaffiltext{2}{Department of Astronomy and Astrophysics,
University of Chicago, Chicago IL 60637, USA} 
\altaffiltext{3}{St. Petersburg State Polytechnical University,  
Politekhnicheskaya 29, 195251 St. Petersburg, Russia.}


\begin{abstract}
A statistical analysis of 
radial distributions 
of  Luminous Red Galaxies (LRGs) 
from the Sloan Digital Sky Survey 
(SDSS DR7)
catalogue within an interval
$0.16  \leq  z  \leq  0.47$ 
is carried out.  We found that 
the radial distribution of $\sim$ 106,000 LRGs incorporates
a few quasi-periodical components relatively to a variable 
$\eta$,  dimensionless line-of-sight comoving distance
calculated for the  $\Lambda$CDM cosmological model.
The most significant peaks of the power spectra 
are obtained for two close periodicities  
corresponding  to 
the spatial comoving 
scales 
$(135 \pm 12)~h^{-1}$~Mpc   and
$(101 \pm 6)~h^{-1}$~Mpc.   
The latter one is dominant and consistent with the 
characteristic scale of the  
baryon acoustic oscillations.
We analyse also the radial distributions of 
two other selected LRG samples:
$\sim 33,400$ bright LRGs ($-23.2 < M \leq -21.8$)
and $\sim 60,300$ all LRGs within a rectangle region 
on the sky,  and  show differences of the 
quasi-periodical features
characteristic for different samples.   
Being confirmed the results would allow 
to give preference 
of  the spatial against temporal models
which could explain 
the quasi-periodicities discussed here. 
As a caveat we show that estimations
of the significance levels of the peaks 
strongly depend on a smoothed radial function
(trend) as well as  
characteristics  of  random  fluctuations.  
%
\end{abstract}



%
\section{Introduction}
\label{s:intro}
It is widely accepted that  Luminous  Red  Galaxies 
(LRGs) are good tracers of the intermediate-  and 
large-scale structures 
of matter in the Universe. 
The procedure of spectroscopic 
identification of  LRGs
from the data of the Sloan Digital Sky Survey 
(SDSS;   e.g.  \citealt{york0}; \citealt{abaz09})
and advantages of its employing for statistical
investigations  were described, e.g. by \citet{eisen01}. 
High intrinsic luminosity of the LRGs and
uniformity of their spectral energy distribution 
allows  to identify     
them at higher redshifts than the main galaxy sample (MGS)
and so to trace a larger volume of the Universe.  

Properties of a spatial  large-scale distribution
of the LRGs were intensively investigated 
over last years 
(e.g. \citealt{eisen05}; 
\citealt{h06}; 
\citealt{percetal07a, percetal07b}; 
\citealt{cg09}; 
\citealt*{gch09};
\citealt{martetal09}; 
\citealt{sccbg09}; 
\citealt{kaz10a, kaz10b}; 
\citealt{labini10}; 
\citealt{percetal10};
\citealt{bl11})
with a     
special interest  to 
a large-scale feature  in their 
two-point correlation function 
and to  associated series of features 
in the  spherically-averaged  power spectrum
P$(k)$ ($h^{-3}$~Mpc$^3$). 
These features  are known to interpret as 
a display of  
the baryon acoustic oscillation (BAO).  
The majority of the cited  papers confirm  the existence
of a significant feature in the distribution of matter, 
although  \citet{labinetal09a}
came to an opposite conclusion
from the analysis of  the
SDSS  Main Galaxy Sample (MGS) --  
data release 7 (DR7;\,     
see also \citealt{labinetal09b}).
  
\citet{kaz10a} 
calculated the two-point correlation
function using a  sample  of  LRGs from the SDSS DR7.    
In the framework of the $\Lambda$CDM cosmological model
the authors have obtained the baryon acoustic peak 
at  (1--2)$\sigma$ significance level and determined
the peak position as $(101.7 \pm 3.0)~h^{-1}$~Mpc
within the cosmological redshift interval $z=0.16-0.36$.
They used the  mock galaxy catalogues  produced by 
the Large Suite of Dark
Matter Simulations  (LasDamas)%
\footnote{http://lss.phy.vanderbilt.edu/lasdamas/mocks.html}
to estimate  a sample variance and systematic errors 
of the  calculations and thus to make their results
more reliable. This result has been 
confirmed recently by  \citet{bl11}  
for an extended redshift range 
$z=0.16-0.44$ with a greater significance 
(3.4~$\sigma$) and a slightly
specified  position, 
$(102.2 \pm 2.8)~h^{-1}$~Mpc,
of the  baryon acoustic peak.

On the other hand,  \citet{rk11} have shown  that 
the radial (line-of-sight) comoving distribution
of absorption-line systems (ALSs) registered in quasar (QSO) spectra 
within the interval $z=0.0-4.3$
reveals a period of $(108 \pm 6)~h^{-1}$~Mpc
or (alternatively) a temporal interval $(350 \pm 20)~h^{-1}$~Myr
for the standard cosmological model. 
The proximity of the scales 
obtained for the spatial LRG distribution, 
e.g. by \citet{kaz10a},
and  the  distribution of ALSs  by \citet{rk11}
suggests to analyse the radial distribution of the same
LRG sample  employing the technics 
used by the latter authors.
     
\begin{figure}[t]   
\includegraphics[width=\columnwidth]{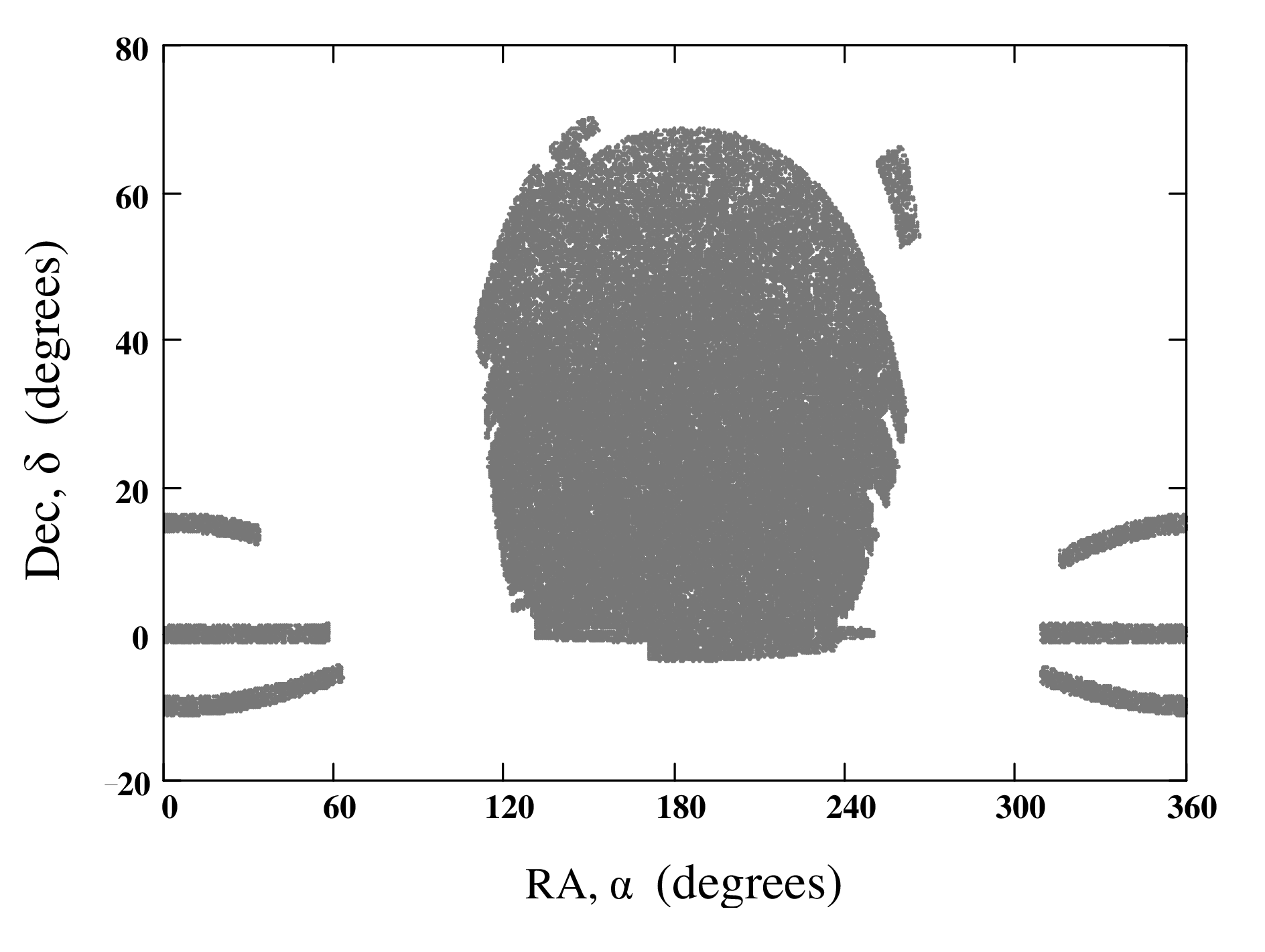}
\caption{
\footnotesize
Angular distribution of  
LRGs  from SDSS DR7 over the sky 
in the Equatorial coordinates;   
all grey coloured  regions comprise    
{\it DR7-Full}  sample of data 
(105 831 LRGs)
represented in Table~1 
by \protect{\citet{kaz10a}};
the data are available
on the World Wide Web (see text). 
}  
\label{sky}
\end{figure}

In this paper we deal with the SDSS DR7 LRG
sample  by \citet{kaz10a}  
presented on the World Wide Web.%
\footnote{http://cosmo.nyu.edu/$\sim$eak306/SDSS-LRG.html}  
The interval of cosmological redshifts considered here, 
$z=0.16$ -- 0.47,  corresponds to the DR7-Full sample
characterized in their  Table~1. 
The SDSS LRG regions on the sky 
are shown in Fig.~\ref{sky}.

The basic quantity of the present study
is a radial distribution  function N$({\eta})$
integrated over  angles  $\alpha$ (right ascension) 
and $\delta$ (declination),
which is an analogue of 
the comoving number density $n(z)$ 
in the redshift space (e.g. \citealt{zehavetal05}); 
$\eta(z)$  is  a dimensionless 
line-of-site  (radial) comoving distance 
between the observer and  galaxies under study,
N$(\eta)$d$\eta$  is a number of LRGs inside 
an interval d$\eta$. 
 
The dimensionless radial comoving distances 
are calculated according to the equation
(e.g. \citealt{h93}, 
\citealt{khs97}, 
\citealt{h99}):
\begin{equation}
  \eta_i =  \eta(z_i)  =   \int_0^{z_i} 
      {1 \over \sqrt{\Omega_{\rm m} (1+z)^3 +
       \Omega_{\rm \Lambda}}} \, \,  {\rm d}z ,
\label{eta}
\end{equation}
where $i$ is a numeration of all LRGs,
$i=1,2, ..., {\rm N}_{\rm tot}$,     
e.g., N$_{\rm tot} = 105 831$ for the 
DR7-Full sample 
by \citet{kaz10a}. 
Following \citet{kaz10a} 
we use   
the dimensionless density parameters
$\Omega_{\rm m}=0.25$ and
$\Omega_{\rm \Lambda}=1-\Omega_{\rm m}=0.75$.   
The corresponding  line-of-sight comoving
distances are\,  
%
%
D$_{{\rm c},i} = c \cdot \eta_i /H_0  = 2998\cdot \eta_i\cdot h^{-1}$~Mpc, 
%
%
where $H_0 = 100~h$~km~s$^{-1}$~Mpc$^{-1}$ is the  
present Hubble constant, $c$ is the speed of light.

We focus our attention
on the search of periodicities 
incorporated
in the radial LRGs distribution 
in relation to the $\eta$-variable.  
We shall treat them as {\it quasi}-periodicities
meaning
limited intervals of line-of-sight 
distances under consideration, 
as well as  variations
of a peak positions and amplitudes 
for  different  samples of  LRGs. 
In Section\ 2 we  examine power spectra   
calculated for two samples of LRGs  
employing a point-like statistical approach:    
the full sample DR7-Full  
and a sample from the rectangle region 
described below.
The point-like approach 
is usually used  
in order to avoid
sensitivity of results
to the procedure of a 
smoothed  distribution  function
(trend)  subtraction  and
to  the related  choice  
of an averaging bin.  
For comparison
in Section\ 3 we discuss power spectra 
calculated in a binning approach 
for the DR7-Bright sample which is
also represented in Table~1  by \citet{kaz10a}.  
In Section\ 4  we use 
two mock galaxy  LasDamas catalogues 
for more conservative estimations of significance levels
of the power spectra,
and outline shortly another 
(widely discussed) way
to estimate the significance of 
the maximal peaks in the power spectra.
Conclusions and  discussions 
of the results 
are given in Section\ 5.
Ambiguity effects of 
a trend  elimination procedure on
results of Fourier spectra analysis 
is shortly outlined 
in Appendix.

\section{Quasi-periodicity of the $\eta$-distribution: \\
point-like approach}
\label{s:point}
 
\begin{figure}[t!]   
\includegraphics[width=1.0\columnwidth]{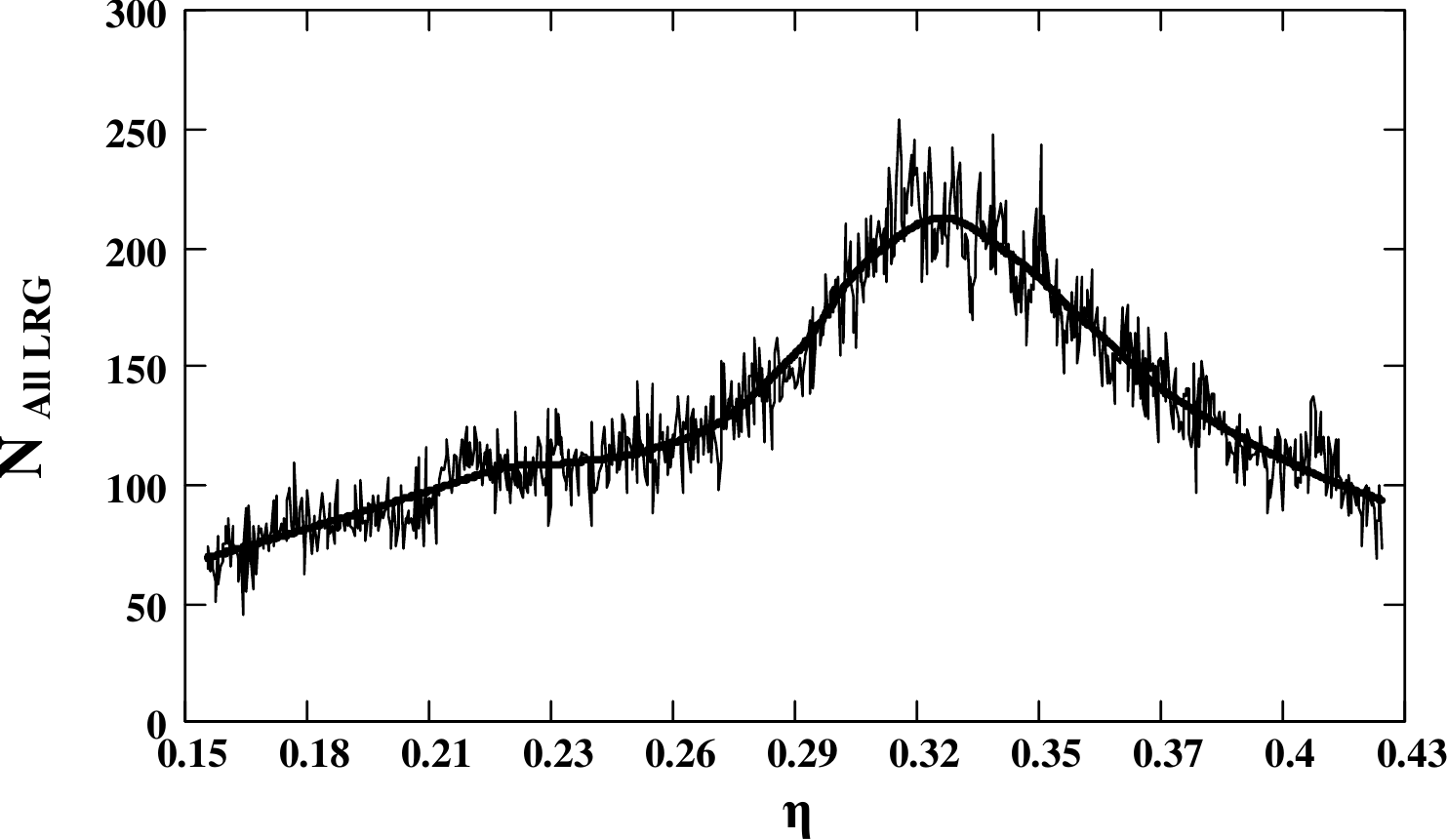}
\caption{
Radial distribution function  
N$(\eta)$ of 105 831  LRGs belonging to 
the grey regions in Fig.~\protect{\ref{sky}}
and  a redshift interval 
$0.16 \leq z \leq  0.47$
or  dimensionless radial distances  
$0.16 \leq \eta \leq 0.43$; 
N$(\eta)$ is calculated  with using
independent bins  $\delta_\eta$ = 0.00033
or  $c \cdot  \delta_\eta/H_0 = 1~h^{-1}$~Mpc; 
thick solid line
displays a trend  N$_{\rm tr}(\eta)$ 
determined as a set of parabolas
(see text).  
}
\label{N(eta)}
\end{figure}

Fig.~\ref{N(eta)} demonstrates  
the radial  
distribution function N$(\eta)$
calculated for independent bins with a narrow width 
$\delta_\eta$ = 0.00033  or\,  
$c \cdot \delta_\eta/H_0 = 1~h^{-1}$~Mpc
at the scale of comoving distances.  
Narrow spike-like variations of  N$(\eta)$  are noticeable 
on the background of the trend
N$_{\rm tr}(\eta)$ drawn by a thick  solid line. 
The function N$_{\rm tr}(\eta)$ 
can be  treated as a smoothed   
function filtering out 
the largest scales       
 (e.g.  \citealt{zehavetal05},  \citealt{kaz10a}).  
We assume that the trend  can be determined 
by selection effects
and/or  large-scale fluctuations in the distribution 
of galaxies  (e.g. \citealt{labini10}).
In Fig.~\ref{N(eta)} the function 
N$_{\rm tr}(\eta)$  is calculated 
by the least-squares method 
using a set of parabolas
as a regression function 
for N$(\eta)$.

Our aim is to reveal weak but quite significant 
quasi-periodical components  
and  separate them from 
the rest determinate part of the 
radial distribution (trend), as well as  fluctuations.   
However, in the case of such a complex trend as 
displayed in Fig.~\ref{N(eta)}
the procedure of the trend  subtraction 
is  ambiguous in principle  and may
bring to wide variations 
of peak positions and  amplitudes in power spectra, as
demonstrated  in Appendix.    
So it is more safely in that case
to employ an out-of-bin (point-like) 
statistical technique,
which is not sensitive to the procedure 
of  trend determination. 

To verify the periodicity 
of the $\eta$-distribution
we calculate a power spectrum 
for the  sequence 
of LRG points $\eta_i \equiv \eta(z_i)$
using the  Rayleigh power:   
(e.g. \citealt{br94}):
\begin{eqnarray}
{\rm P}({\rm m}) &  =  &    
	 {1 \over {\rm N}_{\rm tot}}
                          \left\{ \left[
          \sum_{i=1}^{{\rm N}_{\rm tot}} \cos \left(
            {2 \pi {\rm  m} \eta_i  \over  L_\eta}
              \right) \right]^2   \right.  
\nonumber   \\	      
              	   &   +   & 
             \left.   
	     \left[ \sum_{i=1}^{{\rm N}_{\rm tot}}
             \sin \left(
               {2 \pi {\rm m} \eta_i  \over  L_\eta} 
              \right) \right]^2 \right\},         
\label{P_m}    
\end{eqnarray}
%
m is an integer harmonic number,  
L$_\eta$ = $\eta_{max} - \eta_{min}$
is  an  interval under consideration. 
A periodicity yields a peak in the power spectrum 
${\cal P}$=P(m),   with  
the confidence probability 
or cumulative distribution function
(e.g. \citealt{sc82}) 
\begin{equation}
 \beta = [1 - \exp (- {\cal P})],
\label{beta}
\end{equation}
where $\beta$ is defined   relative to
the hypothesis of  the  Poisson distribution of $\eta_i$.  
This estimation is valid 
for a single independent peak  
at arbitrary m
and  yields the probability of pure noise
generating a power  P(m) 
less than given level ${\cal P}$.    

\begin{figure}[t!]   
\includegraphics[width=\columnwidth]{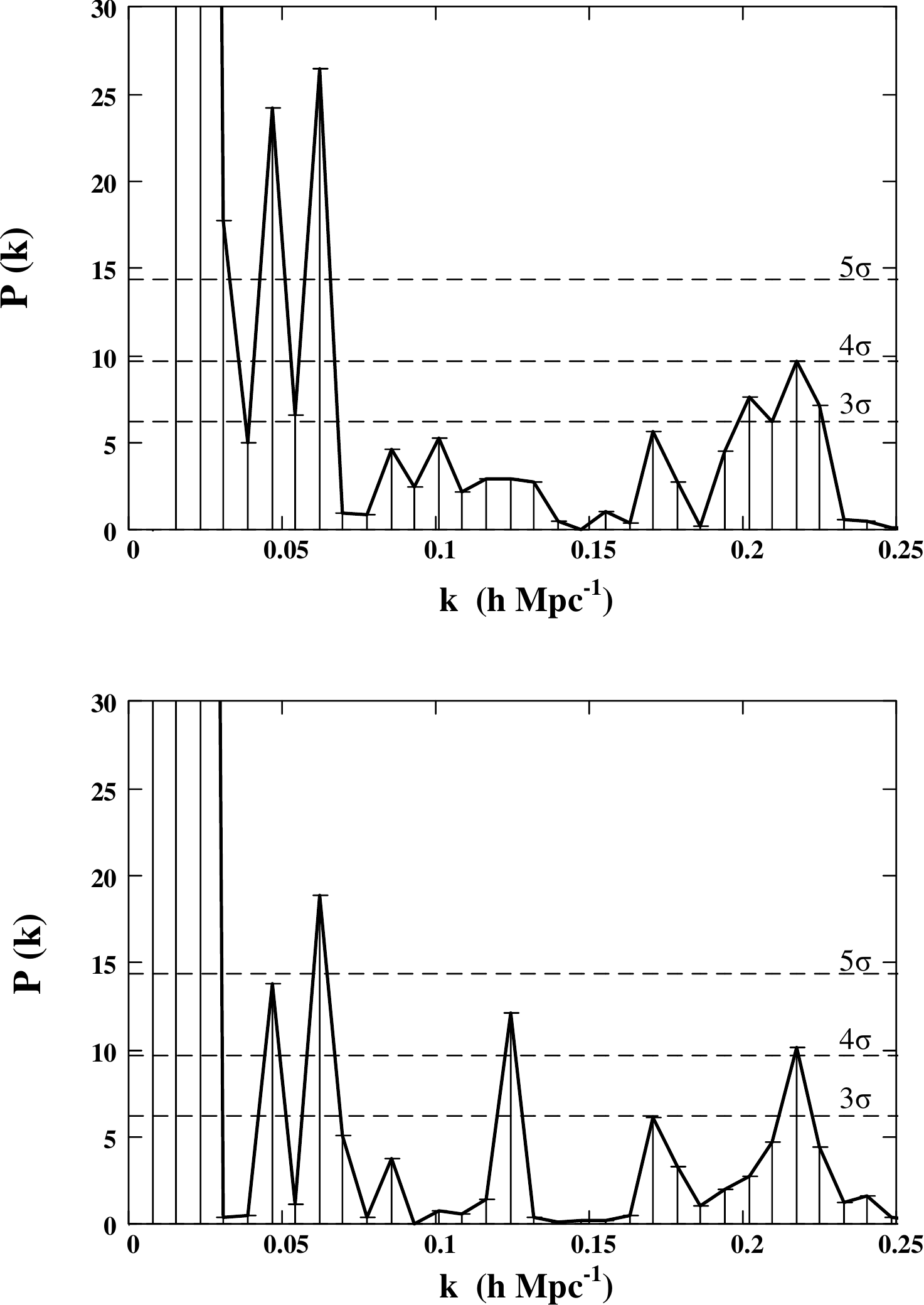}
\caption
{       
Power spectra P($k$) calculated in the point-like approach
according to Eq.~\protect{(\ref{P_m})}
for the redshift interval  $0.16 \leq z \leq  0.47$ 
($0.16 \leq \eta \leq 0.43$), 
$k$ is a wave number; 
the horizontal {\it dash lines} 
specify the significance levels  
3$\sigma$, 4$\sigma$, and 5$\sigma$
estimated with the use of Eq.~\protect{(\ref{beta})}
at fixed levels of the confidence probability
$\beta = 0.998,\  0.99994,\  0.9999994$, 
respectively.
{\it Upper panel:} P($k$) calculated for  all   
LRGs from the  grey regions in Fig.~\protect{\ref{sky}}.       
{\it Lower panel:} P($k$) calculated for LRGs 
from a rectangle region inside  the 
central  oval-like  domain  in Fig.~\protect{\ref{sky}} 
restricted by  intervals of
right ascension, 
$140^\circ \leq \alpha \leq 230^\circ$,   
and  declination, $0 \leq \delta \leq 60^\circ$.  
}
\label{Pk_point}
\end{figure}

Fig.~\ref{Pk_point} represents        
power-spectra P($k$) calculated
with the use of Eq.~(\ref{P_m})   at 
$1 \leq {\rm m} \leq 40$,  where  
the  wave  number  $k$  ($h$~Mpc$^{-1}$)  
substitutes for  m
according  to an expression     
$k = 2 \pi {\rm m}/ {\rm D}_{\rm c}^{\rm L}$,     
where 
${\rm D}_{\rm c}^{\rm L} = c \cdot {\rm L}_\eta/ H_0$  
is the whole comoving interval. 
The upper and lower panels  
correspond to two overlapping samples of LRGs,
both of  them  belong  to
the same interval $0.16 \leq \eta \leq 0.43$
(L$_\eta = 0.27$,\  D$_{\rm c}^{\rm L} = 809.4~h^{-1}$~Mpc).
The upper panel is calculated for
the whole region marked in Fig.~\ref{sky}
comprising the main sample of 105,831  LRGs, 
the lower one --
for a rectangle region  within  the  central  oval-like domain
restricted by the intervals of  right ascension 
$140^\circ \leq \alpha \leq 230^\circ$   
and  declination $0 \leq \delta \leq 60^\circ$. 
The  region  is chosen  
by analogy with  \citet{labini10}
to minimize  possible
effects of the irregular  edges  of  the  central  domain
in Fig.~\ref{sky}.     
The latter sample contains 
60,308 LRGs.          

The most significant peaks of the spectra  P($k$)
are noticeable for two close periodicities
 at $k=0.047$ and $0.062~h$~Mpc$^{-1}$ 
or  harmonic numbers m=6 and m=8.
These values $k$   correspond  to 
dimensionless radial 
comoving scales of  $\eta$\, 
$\Delta_\eta = 0.045 \pm 0.004$  and 
$0.034 \pm 0.002$, 
or   the comoving spatial scales 
$\Delta {\rm D}_{\rm c} = (135 \pm 12)$   and
$(101 \pm 6)~h^{-1}$~Mpc, respectively.   
The latter peak (m=8) is dominant 
for both samples 
with amplitudes well
exceeding  the level 5$\sigma$
estimated  with  Eq.~(\ref{beta}).
While the  first  one (m=6) is somewhat less significant 
being lower than 
the level 5$\sigma$ for the rectangle region.      
  
\begin{figure}[t!]   
\includegraphics[width=\columnwidth]{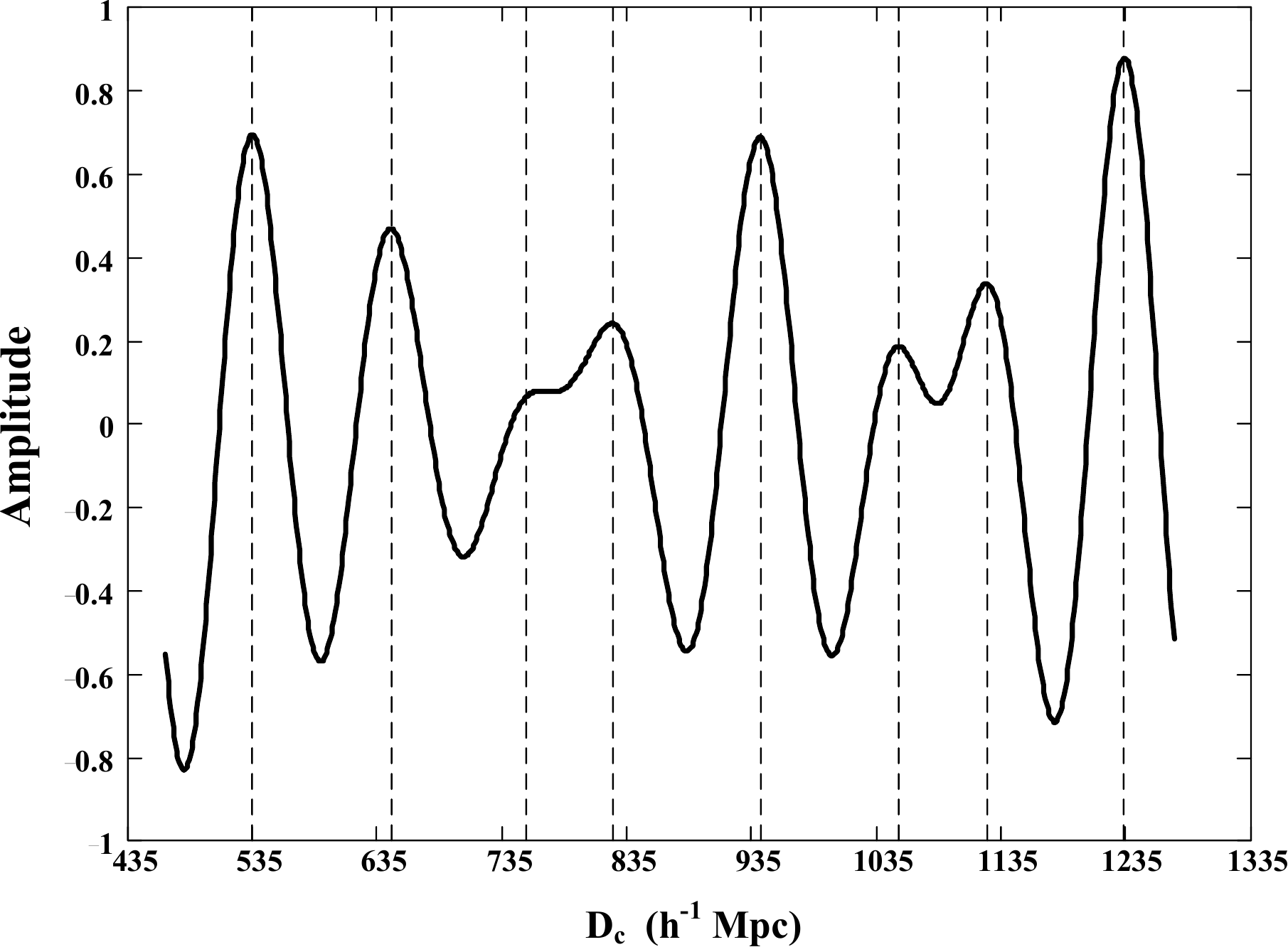}
\caption{
Reciprocal Fourier transformation for five harmonic numbers 
$m$  (see Eq.~\protect{(\ref{P_m})})  from  the
interval $5 \leq m \leq 9$ as a function of  the comoving 
line-of-sight distance D$_{\rm c}$.  
Vertical dash lines indicate
positions of maxima of the reconstructed distribution.
An averaged scale between neighbour dash lines is 
$(100 \pm 8)~h^{-1}$~Mpc. 
}  
\label{m5m9}
\end{figure}

Fig.~\ref{m5m9} demonstrates
a  reciprocal Fourier transform  reconstructed for 
5 harmonics from m=5 till m=9  and  shown  
as a function of 
the comoving  radial distance D$_{\rm c}$. 
The  appropriate direct Fourier transform including amplitudes and phases
of harmonics  is calculated for the sample of all LRGs 
from the  grey  regions in Fig.~\protect{\ref{sky}}.   
One can see that  the pair of close periodical
components m=6 and m=8 
yields a resultant quasi-periodical  fluctuations
of  the radial distribution of LRGs.
An averaged scale of the quasi-periodical  fluctuations 
turns out to be $(100 \pm 8)~h^{-1}$~Mpc,
i.e.  it  is  rather close to the main tone m=8.
 
Less significant peaks are also  visible in 
Fig.~\ref{Pk_point}.  A  wide peak (significance 4$\sigma$)
at $k = 0.217~h$~Mpc$^{-1}$  (m=28), which is  present in the both spectra,
and a narrow peak at $k = 0.124~h$~Mpc$^{-1}$   (m=16)  appearing
only in the  lower  panel.      
The first period corresponds to 
the spatial  scale  
$\Delta {\rm D}_{\rm c}=$ 
$(29 \, \, _{-2}^{+4})~h^{-1}$~Mpc. 
The second one, $(51 \pm 2)~h^{-1}$~Mpc, 
turns out to be very close to the second 
harmonic of the main peak (m=8). 
Note a difference of  the whole sets 
of  spectral features 
for  both   samples under discussion. 

Although the point-like  approach avoids
the procedure of a trend determination 
a steep rise of   both  the 
curves P($k$) at the lowest 
$k  \la 0.05~h$~Mpc$^{-1}$   may  be attributed 
to the effects of  large-scale fluctuations 
incorporated  in the smoothed  $\eta$-dependence 
(see Fig.~\ref{N(eta)}). 
On the other hand,  the long-wave parts  of  the     
spectra are different  for  the  two   
samples  considered  in  Fig.~\ref{Pk_point}.
Such a difference appears, in particular,  
in relative decrease of  the  peak  at  
$k=0.047~h$~Mpc$^{-1}$   
and  cancellation of spectral components 
within an interval of  
$0.03  \la  k \la  0.04~h$~Mpc$^{-1}$ 
in the lower panel.  The difference  could be associated with 
effects of  irregular edges of  the oval-like region 
and a set of strips in  Fig.~\ref{sky}   
excluded  in  the  rectangle  region. 

Our special simulations show  
that  the point-like 
(out-of-bin) approach 
is  more sensitive to the presence of periodicities
in a distribution of  points
than the binning approach  
with  relatively wide  bins.
The latter one  needs  an 
appropriate procedure of a trend subtraction.    
In the case of  simple trends, e.g.
close to linear or parabolic dependencies,  
amplitudes and positions of peaks in the power spectra
calculated  in the binning  mode 
converge to the  values  found in the 
point-like approach  at successive 
reducing of a bin width.  
This convergence 
can be demonstrated, e.g. 
for the narrow independent bins  
$\delta_\eta = 0.00033$    
used in Fig.~\ref{N(eta)}.  
However, we found that  in many cases 
both the convergent procedures  
yield  overestimated significance of peaks
in the power spectra. 
It suggests to re-estimate the significance levels 
of the spectral peaks in Fig.~\ref{Pk_point} 
in a more robust way than it 
can be  produced 
with  using  Eq.~(\ref{beta}).
We  realize  this suggestion in Section~\ref{s:LD}.

\section{Bright LRGs: binning approach}
\label{s:bin}  

In the cases of quite simple  trends   
(linear or parabolic functions)
it is more  relevant  to
use the binning approach and 
a proper procedure of a trend,  N$_{\rm tr}(\eta)$, 
subtraction. For  instance, 
one can calculate so-called 
normalized radial distribution function: 
%
\begin{equation}
{\rm NN}(\eta) =  { {\rm N}(\eta) - {\rm N}_{\rm tr}(\eta)  \over   
                         \sqrt{ {\rm N}_{\rm tr}(\eta) } }.
\label{NN}
\end{equation}
%

Note that instead of the radial distribution function
N$(\eta)$ one can use a comoving number density
$n(\eta)={\rm N}(\eta)/({\rm d}{\rm V}/{\rm d}\eta)$,
where  ${\rm d}{\rm V}/{\rm d}\eta$ is a comoving 
differential volume, which is a simple variation 
of the more conventional value $n(z)$ 
(e.g., \citealt{zehavetal05}, \citealt{kaz10a}). 
This replacement does not change
Eq.~(\ref{NN}) and results of following calculations
because in that case
one should divide into the comoving volume
both numerator and denominator 
(remind that  
$\sigma(n)=\sigma(N)/{\rm V}$, where $\sigma$ is
the mean squared deviation).
   
\begin{figure}[t!]   
\includegraphics[width=\columnwidth]{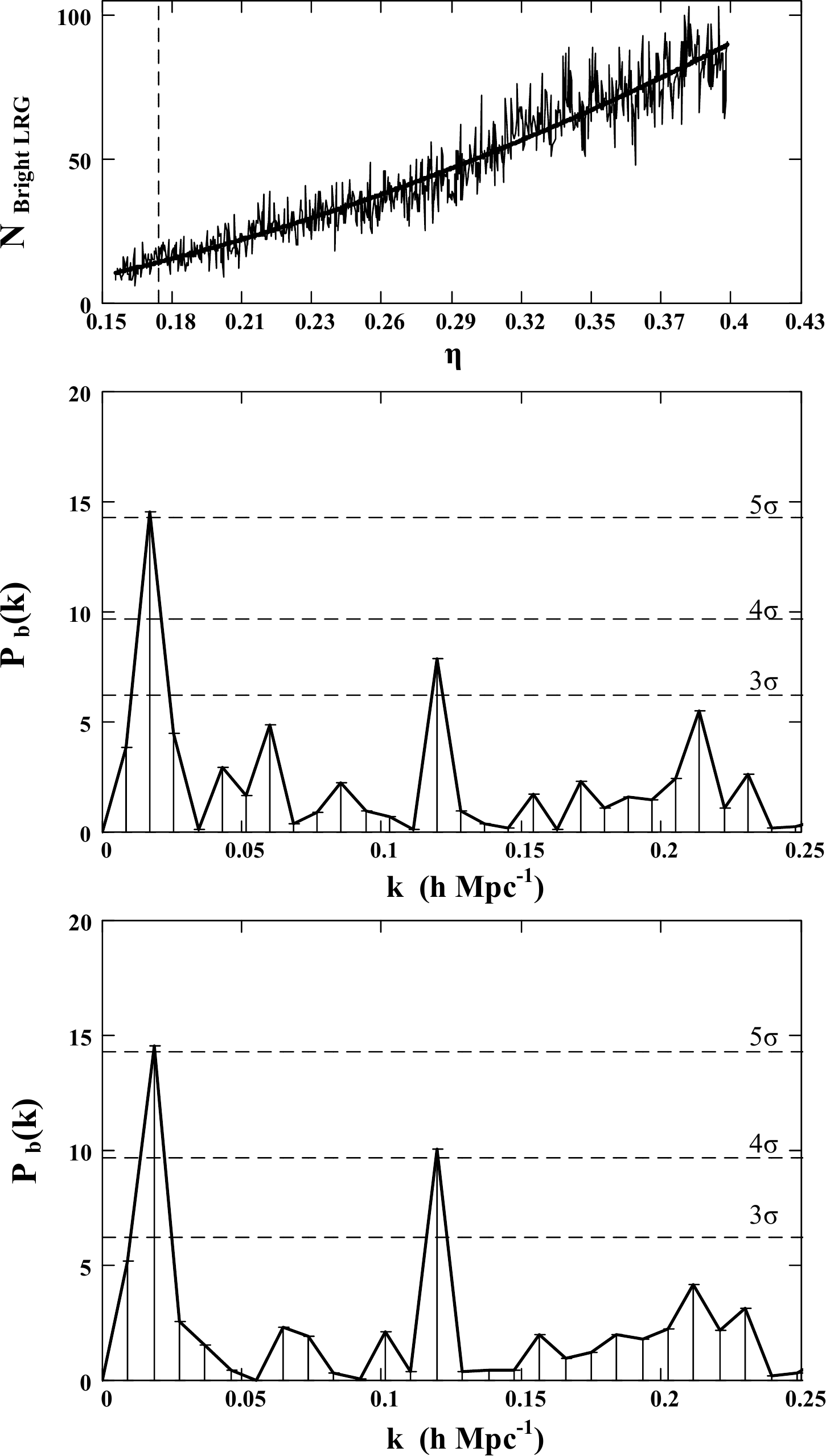}
\caption
{       
{\it Upper panel:}
Radial distribution function N$(\eta)$ 
calculated   
for a sample of  SDSS  ``Bright''  LRGs
($-23.2 < M \leq -21.8$)  within a redshift interval
$0.16 \leq z \leq 0.44$ ($0.155 \leq \eta \leq 0.4$)
indicated as {\it DR7-Bright}  
in Table~1  by \protect{\citet{kaz10a}}.
N$(\eta)$ is calculated  with using
the same independent bins ($\delta_\eta$ = 0.00033)
as in Fig.~\protect{\ref{N(eta)}};
solid  thick   line displays 
a smoothed function (trend) N$_{\rm tr}(\eta)$ drawn as 
2$^{\rm d}$-degree polynomial.
The whole sample contains 33,356 LRGs. 
The vertical dash line shows a shifted low boundary $\eta$
for a reduced sample of the ``Bright'' LRGs.   
{\it Middle panel:}  Power spectrum P$_{\rm b}(k)$
of the function NN$(\eta)$
calculated according to Eq.~\protect{(\ref{Pk_b})}
for the whole sample.
{\it Lower panel:}  Similar power spectrum P$_{\rm b}(k)$
but calculated  for the reduced sample of 32,395 LRGs 
with shifted low boundary:  
$0.172 \leq \eta \leq 0.4$.     
The horizontal {\it dash lines} 
indicate the same  significance levels 
as in Fig.~\protect{(\ref{Pk_point})}  
}
\label{Pk_bin}
\end{figure}

An example of quite simple trend is demonstrated
in the upper panel of Fig.~\ref{Pk_bin}.  
We  explore  the DR7-Bright sample   
indicated in Table~1  by \citet{kaz10a},
which contains  33,356  SDSS  ``Bright''  LRGs
($-23.2 < M \leq -21.8$,\   $M$ is the absolute magnitude)  
within  a redshift interval
$0.16 \leq z \leq 0.44$ ($0.155 \leq \eta \leq 0.40$).
The upper panel in Fig.~\ref{Pk_bin}
displays the function N$(\eta)$  
calculated for the same 
independent bins  ($\delta_\eta = 0.00033$)  
as in Fig.\ \ref{N(eta)}. 
The trend
N$_{\rm tr}(\eta)$ is  calculated as  
a parabolic function by the least-squares method 
and drawn   by  a  thick  solid line.
    
The middle panel in Fig.\ \ref{Pk_bin}
represents the power spectrum  obtained for NN$(\eta)$   
according to the equation: 
%
\begin{eqnarray}
         {\rm P}_{\rm b} (k) & =  & {1 \over {\cal N}_{\rm b}}   
                     \left\{ \left[
          \sum_{j=1}^{{\cal N}_{\rm b}}  {\rm NN}_j  \cos \left(
                  {2 \pi {\rm  m} \eta_j  \over  L_\eta}
              \right) \right]^2  \right. 
\nonumber                                 \\   
	      & +  & 	      
          \left.   \left[ \sum_{j=1}^{{\cal N}_{\rm b}}  {\rm NN}_j
             \sin \left(
                  {2 \pi {\rm  m} \eta_j  \over  L_\eta}
              \right) \right]^2 \right\},
\label{Pk_b}
\end{eqnarray}
%
where  ${\cal N}_{\rm b}$ is a number of bins along $\eta$-axis,
$\eta_j$   
is a location of bin centers, 
$\eta_j \pm \delta_\eta/2$; \,
$j=1,2,..., {\cal N}_{\rm b}$  numerates bins, 
in the  case of the middle panel ${\cal N}_{\rm b} = 736$,\,  
L$_\eta$ = $\eta_{max} - \eta_{min} = 0.245$  
(or ${\rm D}_{\rm c}^{\rm L} = 735~h^{-1}$~Mpc) is
the whole interval.
The power spectrum contains two significant peaks
exceeding the level $3\sigma$ at $k=$ 0.017 and 0.120
$h$~Mpc$^{-1}$. 
The first peak (m=2) corresponds to a half of the  
whole interval and may be interpreted as 
a residual      
large-scale fluctuation which could be
incorporated into the trend N$_{\rm tr}(\eta)$
but it is missed out
by the parabolic approximation. 
The second peak corresponds to a period of
$\Delta {\rm D}_{\rm c} = (52.5 \pm 2)~h^{-1}$~Mpc, 
and may be compared 
with  the peak 
$k=0.124~h$~Mpc$^{-1}$
in the lower panel of Fig.~\ref{Pk_point}.   
  
The lower panel  in  Fig.~\ref{Pk_bin} shows a 
complementary power spectrum   
calculated  in the same way   
but  for a somewhat  diminished 
sample of 32,395 ``Bright'' LRGs within a reduced 
redshift  interval   
$0.18 \leq z \leq 0.44$  ($0.172 \leq \eta \leq 0.40$),
i.e.  lower edge is shifted.
Thus in calculations with  Eq.~(\ref{Pk_b}) 
we use the values 
${\cal N}_{\rm b} = 685$,\,  
L$_\eta$ = $\eta_{max} - \eta_{min} = 0.228$.
Similar to the middle panel
the power spectrum contains two  significant 
peaks at $k=$ 0.018 and 0.120~$h$~Mpc$^{-1}$.  
The first peak,  slightly shifted
in respect of  the peak 
in the middle panel,  implies 
the same interpretation.
The second one  displays  the same period       
as the full sample of  ``Bright'' LRGs but with higher
amplitude  ($\ga 4\sigma$).  
The difference between two spectra may be treated as
an edge effect which provides  better tuning 
of the Fourier component at $k=0.12~h$~Mpc$^{-1}$
in the case of reduced sample.
  
Comparing Figs.~\ref{Pk_bin}  and  ~\ref{Pk_point}
one can notice that the power spectra
are  strongly dependent on a sample of  objects
under investigation.  
In particular, the radial distribution of the bright LRGs
displays  the peak structure of the power spectrum 
which differs from that obtained for the whole sample.

\section{Significance estimations: LasDamas catalogues}
\label{s:LD}  
In this Section we  employ  two  mock 
galaxy LasDamas (LD) catalogs 
``lrgFull-real''  and  ``lrg21p8-real''%
\footnote{http://lss.phy.vanderbilt.edu/lasdamas/mocks/gamma}
(the latter simulates SDSS sample 
of bright LRGs with $M \leq -21.8$)
for complementary estimations of  the significance 
of   spectral peaks  
appearing  in  Figs.~\ref{Pk_point}  
and  \ref{Pk_bin}. 
In both cases we use the data of 80  ``ns'' (North-South) 
realizations.          
Our aim is to take into account possible  
clustering of
the galaxy distribution incorporated in the 
mock LD  catalogs (e.g. \citealt{bw02}, 
see also \citealt{mmetal12})  
in contrast to the  hypothesis of Gaussian fluctuations 
used in  Sections \ref{s:point} and \ref{s:bin}.        
We calculate power spectra 
for all realizations of
both LD catalogs  
and approximate the whole array of spectra 
by the  unified  $\chi^2$-distribution  
applied to 
spectral amplitudes at
each  harmonic number  m 
(wave number $k$) under investigation. 
Quality of the a approximation for 
each m  or $k$  is controlled by 
the Kolmogorov criteria with a tabulated value 
$\lambda_{1-p}$ at fixed significance level 
$p=0.2$ ($\lambda_{0.8}=1.07$). 
 
\begin{figure}[t!]   
\includegraphics[width=\columnwidth]{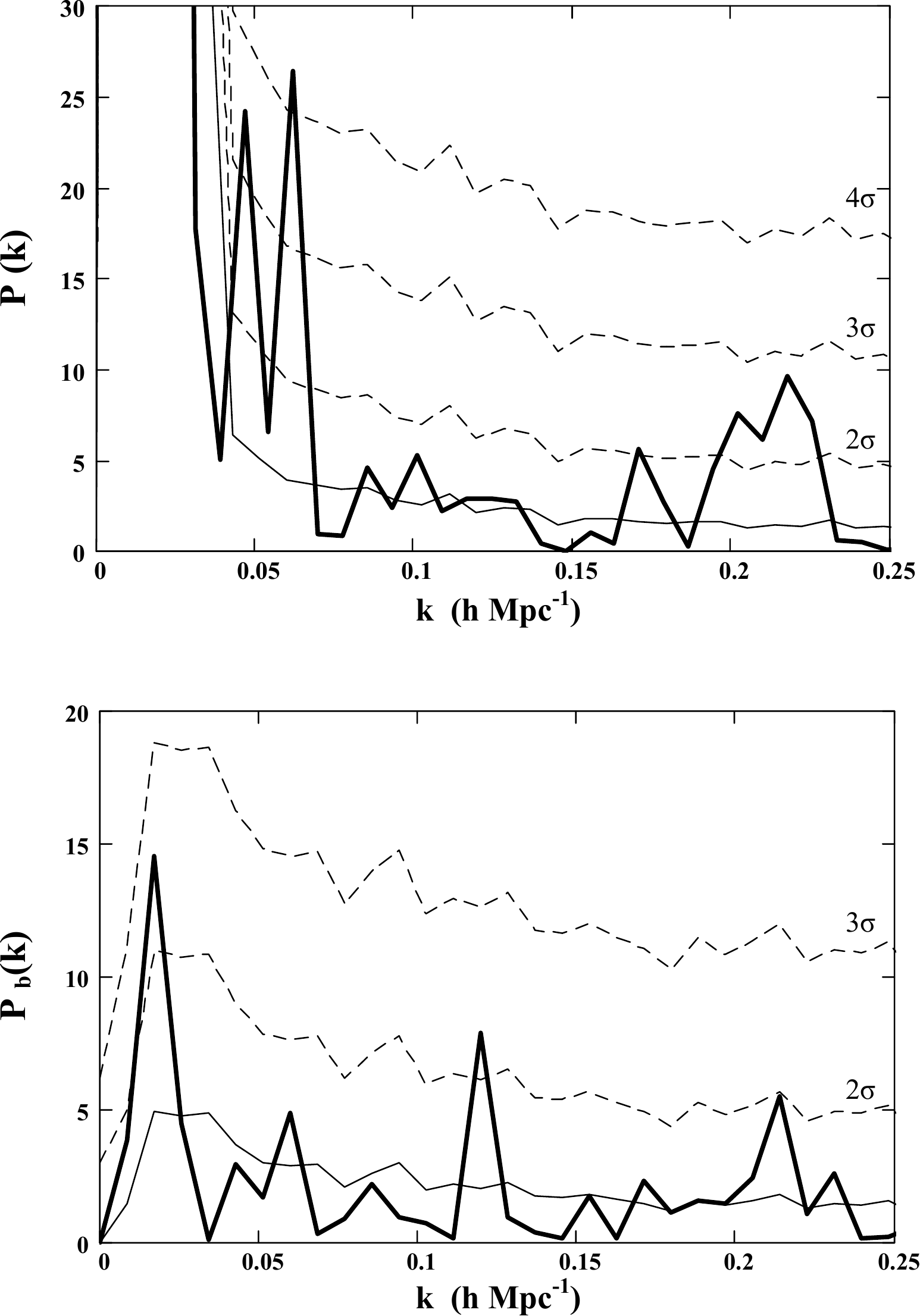}
\caption
{       
Power spectra for two samples of
LRGs  discussed in Sections~\protect{\ref{s:point}} and 
\protect{\ref{s:bin}}  in comparison with appropriate 
significance levels ({\it dash lines}) and  mean
power spectra ({\it thin solid lines}) calculated
on the basis of LasDamas mock catalogs (see text for details).   
{\it Upper panel:} same power spectrum
P$(k)$ as in the  upper panel 
of Fig.~\protect{\ref{Pk_point}}; 
significance levels of $2\sigma$,
$3\sigma$ and $4\sigma$, as well as 
the mean spectral line, are calculated with the 
use of 80 realizations of LasDamas ``lrgFull-real''    
catalog. 
{\it Lower panel:} same power spectrum 
P$_{\rm b}(k)$ as in the  middle panel
of Fig.~\protect{\ref{Pk_bin}}; 
significance levels of $2\sigma$ and
$3\sigma$, as well as 
the mean spectral line, are calculated with the 
use of 80 realizations of LasDamas ``lrg21p8-real''  catalog. 
}
\label{LasDam}
\end{figure}

To  obtain  two sets  of  the power spectra we 
perform 
calculations  of  the radial distributions
of the LD  galaxies  with  using  of  the    
binning approach.  We employ the same bin 
$\delta_\eta$ = 0.00033   
as in  Section \ref{s:bin}.     
As it was  discussed in Section~\ref{s:point} 
the power spectra calculated with   
such a small bin 
in the case of  a simple trend
display insignificant difference 
with the power spectra obtained 
in the point-like  approach. 
This allows one to apply significance levels  
estimated  on the base of  LD data 
to the point-like
statistics used in Section~\ref{s:point}.   

To make statistical properties of the mock    
samples  of  the  ``lrgFull-real''   catalog    
more comparable 
with the  main sample
of LRGs used in  Section~\ref{s:point} 
we implement a reduction of the data,
which reconciles simple radial selection functions (trends)   
N$_{\rm tr}^{\rm LD}(\eta)$
of the  LD  data 
with the complex  trend of the LRG sample.       
The procedure of the reduction 
is performed for   
all realizations of the LD catalog 
within the interval 
$0.16 \leq \eta \leq 0.40$ 
according to the formula:
%
\begin{equation}    
N_{fin} (\eta_j)=N_{in} (\eta_j) \times
{\rm N}_{\rm tr}^{\rm SDSS} (\eta_j)/{\rm N}_{\rm tr}^{\rm LD}(\eta_j);
\label{reduct}
\end{equation}
%
where $\eta_j$  is a centre of  j-th independent bin,
N$_{in} (\eta_j)$  and  N$_{fin} (\eta_j)$
are initial and final  radial distributions 
of  mock   galaxies  
over all independent  bins,  
N$_{\rm tr}^{\rm SDSS} (\eta_j)$ 
is a trend of  the radial  distribution 
calculated for
the main  SDSS  sample, 
N$_{\rm tr}^{\rm LD}(\eta_j)$ 
is a  linear  trend
calculated  for each mock realization. 
For the whole sample of 105,831~LRGs 
we calculate  a  trend  
N$_{\rm tr}^{\rm SDSS} (\eta_j)$ 
as a reciprocal Fourier transformation of the first five
harmonics (m=1, ..., 5)  with their amplitudes and phases
obtained  in the  direct Fourier  transform.
This trend is shown in Fig.~\ref{trend}
of Appendix.  
 
Then  the normalized radial 
distributions  NN$_j$  
can be  determined  by Eq.~(\ref{NN}), 
where
N$_{fin} (\eta_j)$ and  a mean value 
$\overline{\rm N}_{fin}$
stand  for  N$(\eta)$  and N$_{\rm tr}(\eta)$, 
respectively.   The  mean 
$\overline{\rm N}_{fin}$
over  the  interval  
$0.16 \leq \eta \leq 0.40$     
is  used  as the simplest  trend
to  bring  the  binning  approach  of   
Eq.~(\ref{Pk_b} )
closer to the point-like approach  of
Eq.~(\ref{P_m}).   
It  allows  us   to compare statistically a resultant  
sample of the mock power spectra  obtained 
for all realizations of  the  ``lrgFull-real'' catalog
with the spectrum  
in the upper panel of Fig.~\ref{Pk_point}.    

For   the  samples  of  the  ``lrg21p8-real'' mock catalog 
we  also  employ  the  procedure   described   
in  Section~\ref{s:bin}  with  the use of 
N$_{in} (\eta_j)$  and  the parabolic   
trend  N$_{\rm tr}(\eta_j)$   for    
calculations  of   the  values  NN$_j$. 
Following  Eq.~(\ref{Pk_b})
we obtain
a set  of  power spectra  
and  compare  the distributions of 
their amplitudes  
with the power spectrum in the middle
panel of Fig.~\ref{Pk_bin}.   
    
It was found with the use of the Kolmogorov criterion 
that the  $\chi^2$-distribution 
of  peak amplitudes P($k$) is valid for 
a majority (but not  all) of $k$
in the power spectra obtained
for ``lrgFull-real'' and ``lrg21p8-real'' 
realizations.
Specifying the confidence probability $\beta$ 
we calculate 
the significance levels
of P($k$) at each $k=2\pi m/{\rm D}_{\rm c}^{\rm L}$,
determined by all integer  
${\rm m}_{min} \leq  {\rm m}  \leq  40$;
it  is performed  within the 
interval $0.04 \leq k \leq 0.25~h$~Mpc$^{-1}$  (m$_{min} = 5$)
for the whole SDSS LRG sample 
and  at
$0.017 \leq k \leq 0.25~h$~Mpc$^{-1} $  (m$_{min}=2$)
for the ``Bright'' LRGs.

Results of the statistical estimations 
of the  peaks  significance    
for both  SDSS LRG samples 
are represented in Fig.~\ref{LasDam}.
The upper panel displays the same power spectrum as in 
the upper panel of Fig.~\ref{Pk_point} and the lower panel  
corresponds 
to  the middle panel of 
Fig.~\ref{Pk_bin}. Thin solid lines in both panels 
are drawn through arithmetical means of peak amplitudes
calculated for  each   m   (${\rm m}_{min} \leq  {\rm m}  \leq  40$)  
or respective  $k$      
for  the  appropriate  sets of power spectra.
Dashed lines in both panels are drawn
through significance levels $2\sigma$, $3\sigma$, and $4\sigma$
(upper panel) calculated 
as  the amplitudes P$(k)$  
at  all  respective  $k$
matching  
with the confidence probabilities of the  
$\chi^2$-distributions: $\beta=0.95$, $\beta=0.998$,
and $\beta=0.99994$, respectively.       
Let us note that our model estimations
are quite approximate 
due to lack of statistics of the mock catalogs 
and may underestimate
real significance of the peaks. 
  
In the upper panel of Fig.~\ref{LasDam} 
one can see only two peaks  
$k=0.047$  and  0.062~$h$~Mpc$^{-1}$  
exceeding the significance level  $3\sigma$,  
the latter one is higher than  $4\sigma$.
The significance of the third peak $k = 0.217~h$~Mpc$^{-1}$  
hardly approaches to the level $3 \sigma$
in contrast with its level ($4 \sigma$) 
in Fig.~\ref{Pk_point}.  
Three noticeable peaks in the lower panel of 
Fig.~\ref{LasDam} turn out to be less significant 
($ \ga 2 \sigma$). However,  
the growth of the peak at $k=0.12~h$~Mpc$^{-1}$ 
in the lower panel of Fig.~\ref{Pk_bin}   
as well as an appearance of the peak 
at $k \approx 0.12~h$~Mpc$^{-1}$   
in the lower panel of Fig.~\ref{Pk_point}
give an evidence in favour of possible presence
of the additional comoving scale 
$\Delta {\rm D}_{\rm c} = (52.5 \pm 4)~h^{-1}$~Mpc
in the radial distribution of LRGs.
This peak should be confirmed or rejected by further
investigations with extended statistics.

It was marked in Section\ 2 that 
the estimations of 
the confidence probability
according to Eq.~(\ref{beta}) 
imply appearance of a single 
peak  at some wave number $k$ in a power spectrum
with an amplitude ${\cal P}$. 
Following \citet{sc82} (see also \citealt{fef08} and
references therein) one can consider  
a set of many independent wave numbers 
$k_m$;  m$=1,..., N_k$, where 
$N_k$ may be  $\gg 40$ used  in 
Fig.~\ref{Pk_point}, 
and treat any of power peaks P(m)
as a result
of Gaussian noise.
Then one can   
estimate so called
{\it false alarm probability},
$F={\rm Pr}({\rm P}_{max} \geq {\cal P})=1-\beta^{N_k}$,
where $\beta$ is defined in Eq.~(\ref{beta}),
i.e. probability of at least one 
of peaks P$_{max}$          
being equal to 
(or above) a maximal level ${\cal P}$.

Considering 
a set of natural wave numbers 
$k\propto$m for
the whole sample of LRGs 
in Figs.~\ref{Pk_point} and
\ref{LasDam} we regard that  
$N_k$ is close to the theoretical
value ${\cal N}_{\rm b}/2$  
(\citealt{fef08}) 
and exploit also 
the proximity of the point-like and binning approaches 
at the same $\delta_\eta$ = 0.00033. 
One can deal with the normalized power 
amplitude P$_{max}={\cal P}=26.4$
at k$=0.062$  and  use  the formula 
${\cal P}_0=-ln[1-(1-p_0)^{1/N_k}]$ 
(\citealt{sc82},  \citealt{fef08}) ) 
to match a given level of the 
false alarm probability $p_0$ with 
an amplitude ${\cal P}_0$.
In our case  ${\cal N}_{\rm b}=818$,\
$N_k=409$\  and we obtain that the confidence
level $(1-p_0)=0.9999994$ 
(significance $5\sigma$)
corresponds to  ${\cal P}_0=20.3$.
Thus the obtained value
lies between the $5\sigma$-level
in Fig.~\ref{Pk_point} 
($\la 15$) and  the amplitude 
26.4 mentioned above.  
The estimation makes 
the significance of the main peak lower
but not so much lower as 
the $4\sigma$-level
in Fig.~\ref{LasDam}.

Analogous estimations can be carried out 
with the power peak at $k=0.12$ in 
the middle panel of Fig.~\ref{Pk_bin}.
The confidence probability
of the peak 
$1-p_0=0.986$ also becomes lower 
(significance $< 3\sigma$). 
In general, 
the estimations employing the false
alarm probability yield systematically
lower confidence levels
of the maximal power peaks,
but these levels are
higher than the estimations 
on the base of
LD mock catalogs discussed above.   
Therefore we treat the single-peak 
estimations with using Eq.~(\ref{beta}) 
and the LD-estimations as maximal
and minimal boundaries of  
possible confidence levels.    
 
\begin{figure*}[t!]   
\includegraphics[width=2.0\columnwidth]{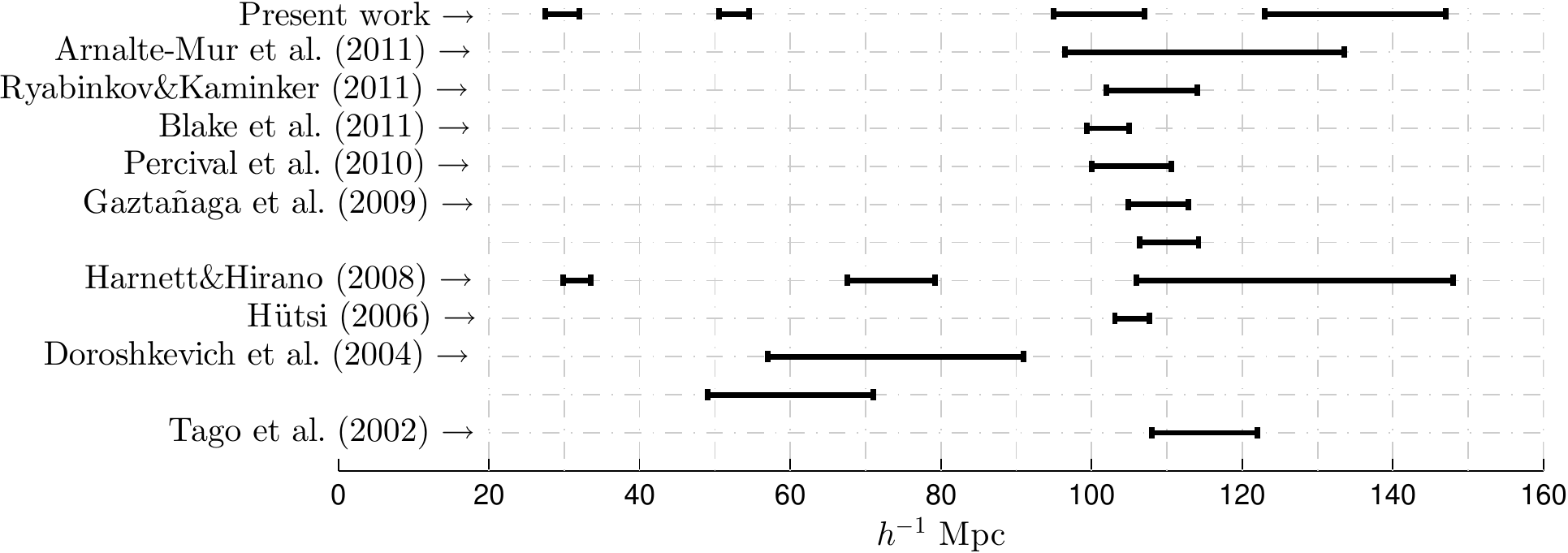}
\caption
{       
Cosmological characteristic scales 
obtained in the last decade by a few groups of authors,
the references are indicated in the left column.    
Horizontal segments depict uncertainties indicated by the authors
except the error bar  attributed to the results of  
\protect{\citet{percetal10}} (at $\Omega_{\rm m}=0.28$,\,  $z=0.2$),
which we estimated on the basis of data given
in the cited paper. The error bar  related to the results of 
\protect{\citet{arnaltetal11}} is also introduced  rather 
arbitrary on basis of their Figure~2.
}
\label{scales}
\end{figure*}
    
\section{Conclusions and discussion}
\label{s:concl}
The main conclusions of the 
statistical analysis  
of radial (line-of-sight) distributions of the LRGs  
within the redshift interval
$0.16  \leq  z  \leq  0.47$ 
or the interval
of dimensionless line-of-sight comoving distance
$0.16 \leq \eta \leq 0.43$
can be summarized as follows:

(1) The radial  distribution  
of $\sim$ 106,000 LRGs incorporates
a few  ($\leq 4$) significant 
quasi-periodical components 
in relation  to the smoothed
function (trend).
The most significant peaks 
of the power spectra 
are displayed for two close periodicities  
corresponding  to the spatial
characteristic scales 
$(135 \pm 12)$  and
$(101 \pm 6)~h^{-1}$~Mpc.   
The latter scale is a dominant with 
significance $\ga 4\sigma$.
This pair of close periodical
components may be  treated as an integrated 
quasi-periodical scale ($100 \pm 8)~h^{-1}$~Mpc. 

(2) Once more appreciable quasi-periodical scale, 
$(29 \, \, _{-2}^{+4})~h^{-1}$~Mpc, arises
in the power spectra as a wide peak  
at a level   $\ga 3 \sigma$.
The peak displays  similar amplitudes
for two  overlapping samples:
the main sample
of 105,831  LRGs
and a sample of  60,308 LRGs 
from the rectangle region (see Sect.~\ref{s:point}).     
Still less reliable scale 
$(52.5 \pm 4.0)~h^{-1}$~Mpc  
appears  for  the  sample  of  LRGs 
from the rectangle-region  
and  for the sample of  ``Bright'' LRGs.
These periodicities must be verified by 
further examinations.

(3) The main  quasi-periodical scale  found  in this work
($(101 \pm 6)~h^{-1}$~Mpc) 
is  consistent with the scale 
$(102.2 \pm 2.8)~h^{-1}$~Mpc 
of the baryon acoustic peak in the two-point
(monopole) correlation function of the LRG distribution 
(\citealt{bl11}). 
These scales are in  agreement also  with  
a period  of  $(108 \pm 6)~h^{-1}$~Mpc
revealed   by \citet{rk11} in the  radial 
distribution of  QSO ALSs
for a  wider redshift  interval   
$0.0 \leq z \leq 4.3$.
Note that the nearness of 
the  periods  $\sim 100 h^{-1}$~Mpc
revealed at  redshifts  $z\la 0.5$
for LRGs  and 
at  $z \leq  4.3$ for
QSO ALSs  argues  for 
spreading of the same large-scale periodicity 
over long interval of the cosmological evolution
(see also \citealt{ddpg11}).

(4) A set of peaks in the power spectra
of Figs.~\ref{Pk_point}, \ref{Pk_bin}
and their variations from sample to sample
probably  evidence in favour of rather spatial   
than temporal nature of the quasi-periodicities
discussed here. Our model simulations performed  
for partly ordered structures of points   
(e.g., \citealt{rk11}) show  that it is
possible  to get a point-like power spectra  similar to 
those represented in Figs.~\ref{Pk_point}
and \ref{Pk_bin}     
performing simulations of
a cloud-like 3D-distribution
of points around  vertices, e.g.  of 
a  face-centered cubic.

(5) On the other hand, we can not  eliminate also
an alternative interpretation which 
may be conventionally denominated as
the temporal one,  i.e. generation of  
some temporal wave processes
in the course of the 
cosmological evolution 
(e.g.  \citealt{m91};  \citealt{ak08};  \citealt{hk10}).
Although, it seems to be more difficult 
to explain a  temporal structure  formed by more-than-one 
periodical  processes.
Note, however, that one temporal interval
$\Delta {\rm T}_{\rm c} = \Delta {\rm D}_c/c =$
$(330 \pm 30)~h^{-1}$~Myr corresponding to the main
quasi-periodicity indicated above could be consistent 
with  the results of \citet{ak08}.  

Let us note that results of our analysis could be
quite sensitive to possible radial incompleteness of the 
DR7-Full sample. It is likely that some fainter LRGs
at larger redshifts were missed by the survey. 
To minimize such a selection \citet{kaz10a} restricted
their analysis by  redshifts $z \sim 0.36$  
corresponding to the quasi-volume-limited sample.
They performed additionally some angular and radial 
weighting procedures and showed that their results
did not variate essentially. Using the same
catalog \citet{bl11} extended
similar analysis to higher redshifts $z=0.44$ and
obtained the close peak position in the  
correlation function. Here we analyse the whole
DR7-Full sample of LRGs up to $z_{max}=0.47$.
However, as it seen in Fig.~\ref{N(eta)}
the distribution function N$(\eta)$ is still
quite representative at
the edge of highest $\eta$.  
All that may evidence in favour
of an assumption that at least the scales
$\sim 100~h^{-1}$~Mpc  are not  distorted 
strongly by the radial-selection effects.         
 
Fig.~\ref{scales}  gives
examples of characteristic scales 
which have been
obtained in the last decade
by a few groups of authors as a result of analysis
of large-scale distributions of galaxies and galaxy clusters. 
A part of data  displayed in Fig.~\ref{scales} 
is taken from Table~1 of \citet{rk11}. 
Additional segments of scales are introduced with using
results of \citet{h06}  on the BAO 
incorporated in 
the power spectrum 
of the SDSS DR4 LRG sample, 
and  \citet{percetal10}   on the 
disclosure of the
BAO in the power spectra
of  both SDSS DR7 samples: LRGs and MGS,
with inclusion of 2dF Galaxy Redshift Survey data.  
We add also the recent results 
by \citet{arnaltetal11}
on the wavelet 
analysis of  acoustic wave features in the 
spatial galaxy distribution.   
  
Fig.~\ref{scales} should be considered only
as an illustration and  it includes rather nonuniform results. 
For instance,  \citet{tagoetal02} yielded a characteristic
period, $(115 \pm 7)~h^{-1}$~Mpc,  of regular spatial oscillations
of the two-point correlation function calculated for
galaxy superclusters 
(see also \citealt{einetal97a,  einetal97b}).   
Two scale intervals  chosen from  Table~1  of  \citet{doretal04}
correspond to the mean separations between walls --
the largest elements of the Large Scale Structure (LSS)
in the galaxy distribution.    
The scale $(74 \pm 17)~h^{-1}$~Mpc was obtained on the basis of radial
(line-of-sight) measurements  with using data of 
SDSS DR1  catalog (at $z \leq 0.14$), 
while the scale $(60 \pm 10)~h^{-1}$~Mpc 
corresponds to  
the Las Campanas Redshift Survey data 
(at $z \leq 0.11$).    
Three  significant periods, $(127 \pm 21)~h^{-1}$~Mpc,     
$(73.4 \pm 5.8)~h^{-1}$~Mpc, and $(31.7 \pm 1.8)~h^{-1}$~Mpc,
were revealed  by  \citet{harhir08} 
in their Fourier analysis of radial galaxy 
distribution on the basis of SDSS DR5 and 2dF GRS data. 
The periodicity found by \citet{rk11}
in the radial distribution 
of QSO ALSs is also included.   

In spite of heterogeneity of the results,
the scales in Fig.~\ref{scales}
may be  formally  subdivided into three groups:
two scales belong  to the interval $(30 \pm 3)~h^{-1}$~Mpc,
four scales belong  to the wider interval 
$(70 \pm 21)~h^{-1}$~Mpc, and the most representative interval
is  $(121 \pm 26)~h^{-1}$~Mpc. Four  periodical components
discussed in the present  paper get into  these intervals;
two most significant periods of our analysis 
belong  to the third group. 
Especially good agreement of the main
scale ,  $(101 \pm 6)~h^{-1}$~Mpc, 
occurs with results obtained by \citet{bl11},
\citet{kaz10a}, \citet{percetal10}, and \citet{h06}
for the BAO scale.  

Actually,  BAOs are discussed in literature 
as a scale of the sound 
horizon at  the recombination epoch
(e.g. \citealt{bg03}; \citealt{percetal07b})
displaying itself 
as a single  feature  in the spatial correlation function 
(e.g. \citealt{eisen05}; \citealt{kaz10a}; \citealt{bl11}). 
This  feature  appears  also  as a series of  
regular variations (oscillations)
imprinted in the power spectrum
calculated for 
the  spatial 3D-distribution of galaxies 
(e.g. \citealt{eih98};  \citealt*{eiht98}; 
\citealt{h06};  \citealt{percetal10}; \citealt{ross12}).
They  principally differ  from the  quasi-regular 
variations of  the radial  (1D)  distributions 
in real space discussed 
by  \citet{rk11}  and   the present paper.
However,  it is not unlikely that  both types 
of variations  could be reconciled.

The  radial distribution variations 
probably bring  out  some spatial 
partly-ordered structure of matter
in the early Universe 
(e.g.  \citealt{einetal97b}; \citealt{einetal11})
characterized by a long-range (or intermediate-range) order.
In particular,  one can admit the possibility  that primordial acoustic perturbations,
responsible for the BAO,  could  carry  traces of a partly ordering 
formed  at  some  early  epochs 
(e.g.  radiation-matter equipartition or
recombination).   In that case  
the  radial (1D)  distributions,
as well as  the two-point  3D-space correlation function,
would display a set of  quasi-periodic features,
which   might  reveal  itself  into   
a complex set  of  features  
(e.g. peaks)  in appropriate power spectra.         
Note  that  possible  presence  of  the second  feature  
in  the two-point  correlation function 
at  a scale approximately double to the 
BAO scale $\sim 200~h^{-1}$~Mpc 
has been  revealed  recently by 
\citet{ross12};  see also \citet{martetal09}.

In any case,     
the existence  of  the large-scale  periodicities,
as well as  the space-ordering  hypothesis, 
needs  further  verifications  based on statistical 
properties of  different cosmological objects over 
wider redshift regions.

\textit{Acknowledgments}
The work has been supported partly
by the RFBR (grant No.\  11-02-01018-a), 
by the State Program ``Leading Scientific 
Schools of Russian Federation'' (grant NSh\ 4035.2012.2),
as well as by
Ministry of Education and Science of Russian Federation 
(contract  No.\ 11.G34.31.0001 and agreement No.\ 8409). 





\appendix
\twocolumn
\section{Effects of trend uncertainties}
   
\begin{figure}[t!]   
\includegraphics[width=\columnwidth]{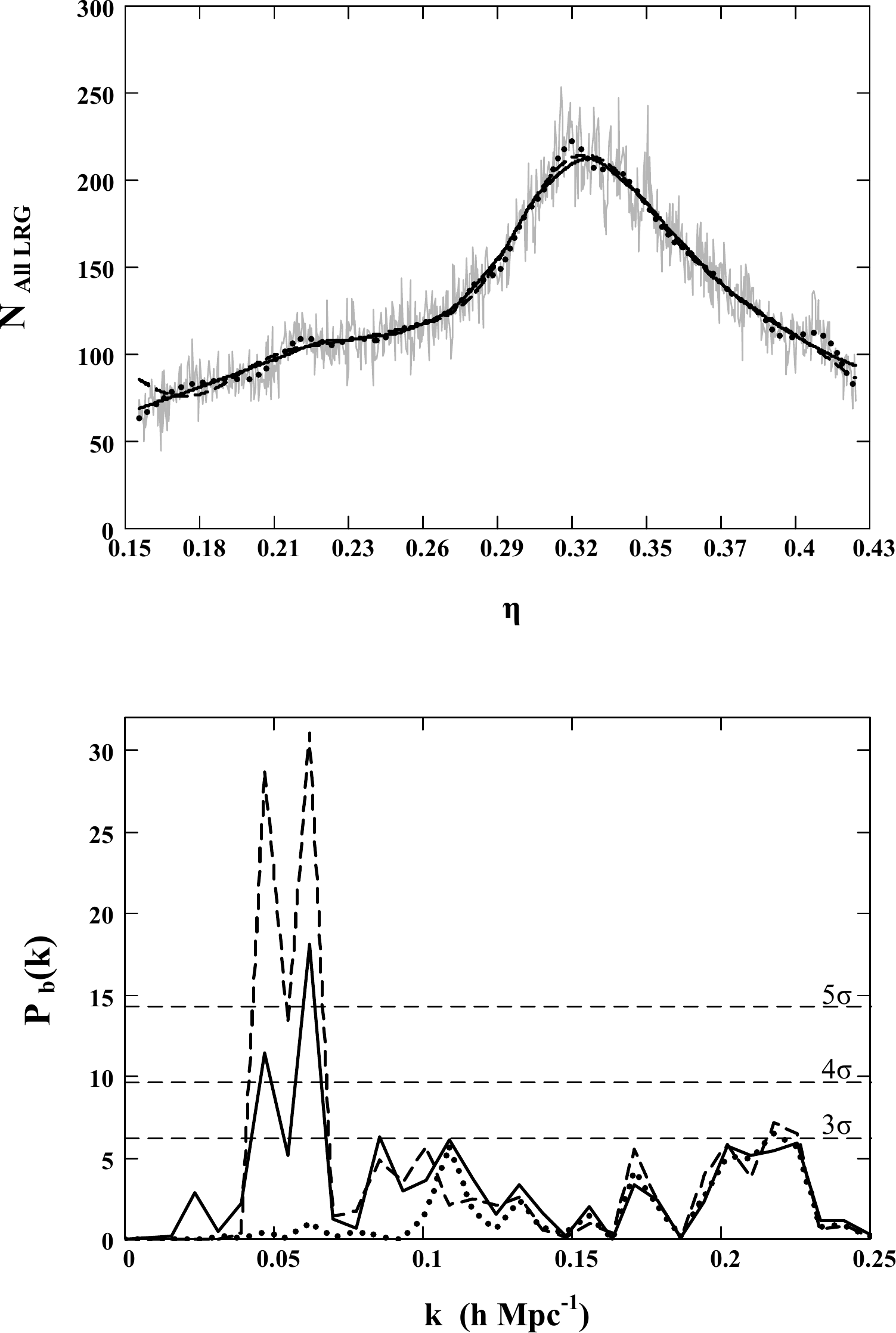}
\caption
{       
Effects of a procedure of  smoothed  function (trend)  
determination 
on  appropriate  power spectra calculated
in the binning approach. 
{\it Upper panel}: 
same radial distribution function  
N$(\eta)$ as in Fig.~\protect{\ref{N(eta)}}
but with three different  trends:
{\it solid} line -- set of parabolas
(same as N$_{\rm tr}(\eta)$ in Fig.~\protect{\ref{N(eta)}}),
{\it dashes} -- reciprocal Fourier transform 
employing  first five harmonics, 
{\it dots}  -- radial selection function obtained  
from a comoving number density $n(z)$ 
of   Kazin et al. (2010a) with  carrying out  a 
spline interpolation.  
{\it Lower panel}:
three power spectra calculated 
with using  Eqs.~\protect{\ref{NN}},  
~\protect{\ref{Pk_b}}  and the trends
displayed 
in the {\it upper} panel by the respective 
type of lines.   
\vspace{150pt}
}
\label{trend}
\end{figure}

Fig.~\ref{trend}  demonstrates  variations 
of the power spectra  obtained  with using
different  procedures  of 
the  radial  smoothed  function (trend)
subtraction  (see Sect.~\ref{s:bin}).
We  examine  105 831~LRGs from DR7-Full sample 
represented  in Table~1 by \citet{kaz10a}.
The upper panel in Fig.~\ref{trend} 
is designed similar to Fig.~\ref{N(eta)}
but  with  using  
three different ways of   
the trend determination.
In all cases
the radial (line-of-sight) 
distribution function N$(\eta)$ is
calculated for independent bins  
$\delta_\eta$ = 0.00033  or\,  
$c \cdot \delta_\eta/H_0 = 1~h^{-1}$~Mpc
providing narrow spike-like variations of 
N$(\eta)$. The first trend function 
N$_{\rm tr}(\eta)$  
is calculated  
by the least-squares method with using 
a set of parabolas (thick {\it solid line})
as a regression function 
for  N$(\eta)$. The second one is calculated
as a sum of  the  first  five reciprocal Fourier harmonics,
m=1, ..., 5 (thick {\it dashed line}), 
with full filtration of all the rest 
harmonics $6 \leq {\rm m} \leq 40$.   
The third trend is calculated with using
the radial selection function $n(z)$ 
by \citet{kaz10a} 
presented on the World Wide Web (see Sect.~\ref{s:intro}).
We rescale $n(z)$ into N$_{\rm tr}(\eta)$  and applicate
a spline interpolation
for  additional smoothing of the trend  function 
({\it dotted line}).
Let us emphasize that  $\chi^2$-criterion does not admit
to prefer one of  three trends although the
results of power spectrum calculations 
for each of  them are  qualitatively different. 

The lower panel displays the respective power spectra 
calculated with  Eqs.~(\ref{NN})  and  (\ref{Pk_b})  and
depicted by the same type of lines. One can see how
the two most significant peaks at  $k=0.062~h$~Mpc$^{-1}$  (m=8) 
and  $k=0.047~h$~Mpc$^{-1}$  (m=6) strongly depend on a trend
obtained in one of  the  ways  indicated above:
the  features  may  variate   
from strong double peak  ({\it dashed line})
up to  its  cancellation ({\it dotted line}). 
Whereas   for $k \ga 0.1$, i.e. for shorter  waves,
the effects of different trends are  less  essential.   
One  can  notice  that  in  the  upper panel 
the spline interpolation of the trend 
(dots) displays 
long-wave oscillations and thus compensates
the  corresponding Fourier harmonics.     
To some degree  such a compensation may
occur  for different  trends subject to their
sufficient  complexity.  
The  oscillations inherent  in  trends 
can  have  also as physical
as  nonphysical  origin  and 
in their  turn need  
special
statistical considerations.

The  indicated uncertainties  make  
the binning approach, including
a procedure of the trend subtraction,
quite ambiguous at least in  
the cases of complex trends. 
Therefore,     
one needs a  verification 
by a  few  statistical   tests
additional  to   
the binning approach  discussed in Sect.~\ref{s:bin}, 
including  the point-like  treatment, 
to make more robust  conclusions
concerning  reality  of   quasi-periodicities
in the distribution of matter.



\end{document}